\documentstyle[12pt]{article}
\renewcommand{\i}{\imath}
\renewcommand{\j}{\jmath}
\newcommand{\bi}{{\bar\i}}
\newcommand{\bj}{{\bar\j}}
\newcommand{\bk}{{\bar k}}
\renewcommand{\=}[1]{\bar{#1}}
\newcommand{\sect}[1]{\section{#1}\setcounter{equation}{0}}

\begin{document}

\bigskip
\hskip 5in\vbox{\baselineskip12pt
\hbox{NSF-ITP-00-111}
\hbox{hep-th/yymmnnn}}
\bigskip\bigskip

\centerline{\Large Supersymmetric Three-Form Flux}
\centerline{\Large Perturbations on $AdS_5$}
\bigskip
\centerline{\bf Mariana Gra\~na} 
\medskip
\centerline{Department of Physics}
\centerline{University of California}
\centerline{Santa Barbara, CA 93106}
\centerline{\it mariana@physics.ucsb.edu}
\bigskip
\centerline{\bf Joseph Polchinski} 
\medskip
\centerline{Institute for Theoretical Physics}
\centerline{University of California}
\centerline{Santa Barbara, CA\ \ 93106-4030}
\centerline{\it joep@itp.ucsb.edu}

\begin{abstract}

We consider warped type IIB supergravity solutions with three-form
flux and ${\cal N}=1$ supersymmetry, which arise as the supergravity
duals of confining gauge theories.  We first work in a perturbation
expansion around $AdS_5
\times S^5$, as in the work of Polchinski and Strassler, and from the
${\cal N}=1$ conditions and the Bianchi identities recover their
first-order solution generalized to an arbitrary  ${\cal N}=1$
superpotential.  We find the second order dilaton and axion by the same
means. We also find a simple family of exact solutions, which can be
obtained from solutions found by Becker and Becker, and which
includes the recent Klebanov--Strassler solution.

\end{abstract}

\newpage
\baselineskip=17pt

\sect{Introduction}

The duality proposed by Maldacena \cite{Maldacena} between type IIB string
theory in $AdS_5 \times S^5$ and ${\cal N}=4$  Yang Mills in four dimensions
has
been an arena for amazing theoretical advances in the past few years. This
duality applies to a
conformal four dimensional theory, and so as it stands it
cannot shed light on
confining gauge theories such as QCD\@.  However, from it one can
deduce dualities involving systems with less supersymmetry, including
confining
gauge theories.

In addition to their relation to gauge theories, 
deformations of $AdS_5$ with reduced supersymmetry are of interest for their
connection to the proposal of Randall and Sundrum~\cite{RS1, RS2}.  These
authors suggested that the hierarchy problem could be solved in a higher
dimensional space with a large warp factor.  Such warped spaces can be
realized in string theory in various ways, most directly by bringing
together $N$ D3-branes on a Calabi-Yau manifold to produce a region that
is locally $AdS_5 \times S^5$~\cite{Verlinde}.  In the simplest case
there is ${\cal N}=4$ supersymmetry, but one is ultimately interested in
at most ${\cal N}=1$.

One means of reducing the supersymmetry is by perturbing the Hamiltonian,
which corresponds to perturbing the boundary conditions on $AdS_5 \times S^5$
\cite{witads, GKP}.  The breaking of ${\cal N} = 4$ to ${\cal N} = 1$ by mass
terms has been studied from various points of view~\cite{VafaWitten,
DonagiWitten, Strassler, Dorey, DoreyKumar, gppz}.
The supergravity dual to this perturbed theory appears to contain a naked
singularity, but in a recent paper \cite{PS}
(henceforth PS), Polchinski and Strassler showed that this singularity is
actually replaced by an expanded brane source, so the theory is tractable.

The supersymmetry can also be be reduced by placing D3-branes at a singular
point, such as an orbifold or conifold point~\cite{
kacsil,kehag,klebwit,morpless}, rather than a
regular point.  To break the conformal invariance it is necessary in addition
to introduce fractional branes~\cite{gubkleb,klebnek}.  Again the
supergravity dual appears at first to be singular~\cite{klebtsey}, but recent
work by Klebanov and Strassler (henceforth KS) has shown that the true
solution is nonsingular~\cite{klebstrass}.  Somewhat surprisingly,
explicit branes are not involved; rather, the distinctive feature of the
nonsingular solution is a reduced (spontaneously broken) symmetry.

In this paper we explore warped IIB supergravity
solutions with unbroken ${\cal N} = 1$ supersymmetry.  Our initial goal is to 
understand the supersymmetry of the PS solution.  The ${\cal N}=1$ mass
in the gauge theory is dual to a three-form flux in
the supergravity solution; we work in the same
approximation as in PS, treat this flux as a perturbation.

In section~2 we review the type IIB supergravity fields and supersymmetry
variations, and the zeroth order IIB solution that corresponds to the Coulomb
branch of the ${\cal N}=4$ gauge theory.  In section 3 we first solve the
conditions for unbroken ${\cal N}=1$ supersymmetry at first order
around a general Coulomb branch solution.  We then
impose the Bianchi identities and find that the general solution is
characterized by one holomorphic function $\phi$ and
one harmonic function $\psi$.  We verify that this general solution to the
supersymmetry and Bianchi conditions also satisfies the equations of motion.
The holomorphic function corresponds to an arbitrary ${\cal N}=1$
superpotential, while the harmonic function corresponds a higher dimension
perturbation.

The KS solution also involves a three-form flux, but this cannot be regarded
as a perturbation.  In section~4 we note a simple class of exact solutions,
which includes the KS solution and its ${\cal N}=2$ version~\cite{joemm}.
This class is actually a special case of a class of M/F theory solutions found
by Becker and Becker~\cite{beck^2,GVW,DGS}.

In section~5 we make concluding remarks.  An appendix contains various
extensions of the work in section~3: the second order solutions for the
dilaton and axion (which can be obtained easily because they decouple from
the other second order perturbations); a simple particular solution; and, a
verification that the solution obtained here agrees with that found in PS
(in particular, that the normalizable solution is determined in terms of the
warp factor).

\sect{Review of IIB supergravity}

\subsection{Fields and variations}

The massless bosonic fields of the type IIB superstring theory consist of
the
dilaton $\Phi$, the metric tensor $g_{MN}$ and the antisymmetric 2-tensor
$B_{MN}$
in
the NS-NS sector, and the axion $C$, the 2-form potential $C_{MN}$ and the
four-form field $C_{MNPQ}$ with self-dual five-form field strength in the
R-R sector. Their fermionic superpartners are a complex Weyl
gravitino $\psi_{M}$ ($\gamma^{11}\psi_{M}=-\psi_{M}$) and a complex Weyl
dilatino $\lambda$ ($\gamma^{11}\lambda=\lambda$). The theory has $\mathcal
{N}$=2 supersymmetry generated by two supercharges of the same chirality.
%
%
The
two scalars can be combined into a complex field $\tau = C+
ie^{-\phi}$ that
parameterizes the $SL\left(2,\bf R\right)/ U\left(1\right)$ coset space.

We want to find 
background that preserve some supersymmetry.  Assuming that the
background
fermi fields vanish, we have to find a combination of the bosonic fields
such that the supersymmetry variation of the fermionic fields is zero.
The equations for the variation of the dilatino and gravitino have been
found in
\cite{Schwartz}, whose conventions we use.  We use subindices
$M,N,...=0,...,9$;
$\mu,\nu = 0,1,2,3$ and $m,n,...=4,..,9$
\begin{eqnarray}
\delta\lambda &=& \frac{i}{\kappa} \gamma^{M}P_{M}\varepsilon^*-
\frac{i}{24}\gamma^{MNP}G_{MNP} \varepsilon\ ,
\label{dilatino}
\\
\delta\psi_{M} &=& \frac{1}{\kappa}(D_{M} -{1\over2}i Q_M)
\varepsilon + \frac{i}{480}
\gamma^{M_{1}...M_{5}}
F_{M_{1}...M_{5}}\varepsilon+\frac{1}{96}\left(\gamma_{M}^{PQR}G_{PQR}-
9\gamma^{PQ}G_{MPQ} \right) \varepsilon^{*}\, .\nonumber\\
\label{gravitino}
\end{eqnarray}
Here
\begin{eqnarray}
P_{M} &=& f^2 \partial_M B\ ,
\quad Q_{M}= f^2 {\rm Im}(B \partial_M B^*) \ ,\label{PQ}
\\[2pt]
B &=& \frac{1+i\tau}{1-i\tau}
\ ,\quad f^{-2} = 1 - BB^*
\ .
\end{eqnarray}
The supersymmetry parameter
$\varepsilon$ is a complex Weyl
spinor
$(\gamma^{11}\varepsilon=-\varepsilon)$, and $D_M$ is the covariant
derivative with respect to the metric $g_{MN}$.
The field strengths are
\begin{eqnarray}
G_{(3)} &=& f(F_{(3)}-BF_{(3)}^*)\ ,\quad
F_{(3)} = dA_{(2)} \ ,\\
F_{(5)}&=& dA_{(4)}-{\kappa\over 8}\,{\rm Im}( A_{(2)}\wedge
F_{(3)}^*)\ .
\label{defpotth}
\end{eqnarray} 
with $A_{(2)} = C_{(2)} + i B_{(2)}$ complex and $A_{(4)}$ real.

\subsection{Black 3-brane background}

For any distribution of D3-branes aligned along the $\mu$-directions, the
background is
\begin{eqnarray}
ds^{2} &=& Z^{-1/2}\eta_{\mu\nu}dx^{\mu}dx^{\nu}+ Z^{1/2} dx^{m}dx^{m}\ ,
\label{metricZ}\\
F_{\mu\nu\lambda\rho
m} &=& \frac{1}{4\kappa Z}\epsilon_{[\mu\nu\lambda\rho}\partial_{m]}Z\
,\quad F_{mnpqr}= -\frac{1}{4\kappa Z}\epsilon_{mnpqrs}\,\partial^{s}Z\ ,
\label{F5}
\end{eqnarray}
with $\tau$ constant and $G_{(3)} = 0$.  For $N$ D3-branes at $x^m = 0$,
\begin{equation}
Z= \frac{R^{4}}{r^{4}}\ ,\quad R^4 = 4\pi g N \alpha'^2
\label{Z}
\end{equation}
where $r^2 = x^m x^m$; the spacetime is then $AdS_5 \times S^5$.\footnote
{Note that $\kappa$ and $g$ are related, $2\kappa^2 = (2\pi)^7 \alpha'^4
g^2$.}
 More generally, $Z$ is any function of the $x^m$, and
\begin{equation}
-\partial_m \partial_m Z(x^m) = (2\pi)^4 \alpha'^2 g \rho_3(x^m)
\label{3den}
\end{equation}
where $\rho_3(x^m)$ is the density of D3-branes in the transverse space.

For this background the dilatino equation~(\ref{dilatino}) is
automatically satisfied. The gravitino variations~(\ref{gravitino}) in
this background are
\begin{eqnarray}
\kappa \delta \psi_\mu &=& \partial_\mu \varepsilon -
\frac{1}{8}
\gamma_\mu
\gamma_w (1 - \Gamma^4) \varepsilon \ ,
\\
\kappa \delta \psi_m &=& \partial_m \varepsilon +
\frac{1}{8}
\varepsilon
w_m - \frac{1}{8} \gamma_w \gamma_m (1 - \Gamma^4) 
\varepsilon \ ,
\end{eqnarray}
where $\Gamma^4 = i\gamma^{0123}$ is the four-dimensional chirality
and we have defined
\begin{equation}
w_m = \partial_m \ln Z\ ,\quad\gamma_w = \gamma^m w_m \ .
\end{equation}
The spin connection has been calculated for tangent space axes $\hat M$
parallel to the Cartesian coordinate axes $M$. 
The Poincar\' e supersymmetries are independent of $x^\mu$ and so the
vanishing of $\delta \psi_\mu$ implies that
\begin{equation}
\Gamma^4 \varepsilon = \varepsilon\ .
\end{equation}
The vanishing of $\delta \psi_m$ then implies that
\begin{equation}
\varepsilon = Z^{-1/8} \eta\ .
\end{equation}
where $\eta$ is any constant spinor such that $\Gamma^4 \eta =
\eta$.  We can decompose 
\begin{equation}
\eta = \zeta \otimes \chi
\end{equation}
where $\zeta$ is any four-dimensional anticommuting spinor with positive
$\Gamma^4$-chirality and $\chi$ is any six-dimensional commuting spinor
with negative $\Gamma^6$-chirality.
There are two such $\zeta$ and four such $\chi$, giving eight complex or
16 real supersymmetries.

\sect{Perturbative solutions}

\subsection{Supersymmetry conditions}

The PS solution~\cite{PS} was obtained in a perturbation
expansion in powers of $G_{(3)}$ around the D3-brane
solution~(\ref{metricZ},\ref{F5}).  This expansion was justified in PS
because it was found, in large regions of parameter space, that the
D3-brane density dominated the density of the 5-branes that appeared in
the resolution of the naked singularity.
We can expand the supersymmetry
parameter in the same way,
\begin{equation}
\varepsilon = \varepsilon_0 + \varepsilon_1 + \ldots\ ,
\end{equation}
where as above
\begin{equation}
\varepsilon_0 = Z^{-1/8} \eta = Z^{-1/8} \zeta \otimes \chi\ .
\end{equation}
We are looking for solutions with $D=4$, ${\cal N} = 1$ supersymmetry so
only one choice of the spinor $\chi$ is kept.\footnote{
IIB backgrounds with ${\cal N} = 1$ supersymmetry were studied in
refs.~\cite{candelas,dewit}, but with the assumption
that the transverse dimensions are compact and without brane or other
sources.}

It is convenient to adopt complex coordinates $z^{i}$ for the 6-dimensional
part
of the metric:
\begin{equation}
z^1=\frac{x^4+ix^7}{\sqrt2}\ ,\quad z^2=\frac{x^5+ix^8}{\sqrt2}\ ,
\quad z^3=\frac{x^6+ix^9}{\sqrt2}\ .
\label{zdef}
\end{equation}
In these coordinates, the six-dimensional part of the metric
(\ref{metricZ})
is
\begin{equation}
g_{i\bar\j} = Z^{1/2}\delta_{i\bar\j}\ .
\label{metriccomplex}
\end{equation}
By an $SO(6)$ rotation we can choose $\eta$ such that
\begin{equation}
\gamma_{i} \varepsilon_0 = \gamma^{\bar\i}
\varepsilon_{0}=0\ .
\label{gammaepsilon}
\end{equation}

We now expand the supersymmetry equations in powers of $G_{(3)}$.  In the
PS solution, the dilaton, metric, and five-form receive no first order
correction so the first order dilatino equation is just
\begin{equation}
G\varepsilon_{0}=0
\label{dilatino1}
\end{equation}
where $G=G_{MNP}\gamma^{MNP}=G_{mnp}\gamma^{mnp}$.  Expanding in complex
coordinates and using the property~(\ref{gammaepsilon}), this is
\begin{equation}
0 = G_{ijk} \gamma^{ijk}\varepsilon_{0} + 3G_{\bar\i jk} \gamma^{\bar\i jk}
\varepsilon_{0}
= G_{ijk} \gamma^{ijk}\varepsilon_{0} + 6 G^j{}_{ jk} \gamma^{k}
\varepsilon_{0}
\label{dilatino2}
\end{equation}
The spinors in the two terms are independent, so
\begin{equation}
G_{ijk}=G^j{}_{ jk}=0\ .
\label{Gijk}
\end{equation}

The first order term
in $\delta\psi_\mu = 0$ is
\begin{equation}
-\frac{1}{8\kappa} \gamma_\mu
\gamma_w (1 - \Gamma^4) \varepsilon_1 
+\frac{1}{96} \gamma_\mu G \varepsilon_0^{*}
= 0\ .
\end{equation}
From the structure of the equations we can assume that $\varepsilon_1$
(like $\varepsilon_0^*$) has the opposite chirality to $\varepsilon_0$,
namely $\Gamma^4 = -1$,\footnote
{A term of the same chirality would have to be proportional to
$\varepsilon_0$ and so can be absorbed in the latter.}
 and so this determines 
\begin{equation}
\varepsilon_1 
=\frac{\kappa}{24 S^2} \gamma_w G \varepsilon_0^{*}\ .
\end{equation}
We have defined
\begin{equation}
S^2 = \gamma_w \gamma_w = Z^{-1/2} w_m w_m \ .
\end{equation}

The first order term in $\delta\psi_m = 0$ becomes
\begin{equation}
\partial_m \xi
- \frac{1}{2} Z^{-1/2} G \gamma_m
\eta^* = 0\ ,
\end{equation}
where $\xi = Z^{-1/2} S^{-2} \gamma_w G \eta^*$.
We have used the identity
\begin{equation}
\gamma_{M}^{PQR}G_{PQR}-
9\gamma^{PQ}G_{MPQ} = - 2 G \gamma_M - \gamma_M G\ .
\end{equation}  
For $m = \bar\i$, the property (\ref{gammaepsilon}) immediately gives
\begin{equation}
\partial_{\bar\i} \xi = 0
\end{equation}
so that $\xi$ is a spinor holomorphic in the $z^i$ and
\begin{equation}
G \eta^* = Z^{1/2} \gamma_w \xi(z)\ . \label{Geta}
\end{equation}
The final, $m = i$ equation, then becomes
\begin{equation}
G \gamma_i \eta^* = 2 Z^{1/2} \partial_i \xi(z)\ . \label{Gieta}
\end{equation}
This can also be written with $i \to m$, as the $\bar\i$ components hold
trivially. 

We now wish to expand eqs.~(\ref{Geta},\ref{Gieta}) in terms of the
components of $G$.  The most general holomorphic spinor $\xi$ of the
correct chirality is
\begin{equation}
\xi(z)=\alpha \eta^* + \frac {1}{2} \beta_{\bi\bj}
\gamma^{\bi\bj}
\eta^*\ ,
\label{xidescomp}
\end{equation}
where $\alpha(z)$ and
\begin{equation}
\hat\beta_{\bi\bj}(z) = Z^{-1/2} \beta_{\bi\bj}
\end{equation}
are holomorphic.  The factor of $Z^{-1/2}$ arises because it is the
$\gamma^{\hat m}$, with tangent space index, that are matrices with
constant coefficients, whereas
\begin{equation}
\gamma^m = Z^{-1/4}\gamma^{\hat m}  
\end{equation}
are position-dependent.
Also,
\begin{eqnarray}
G \eta^* &=& G_{\bi\bj\bk} \gamma^{\bi\bj\bk} \eta^* + 3
G_{\bi\bj k} \gamma^{\bi\bj k} 
\eta^* \nonumber\\
&=& G_{\bi\bj\bk} \gamma^{\bi\bj\bk} \eta^* + 6
G^{\bi}{}_{\bi\bj} \gamma^{\bj} \eta^* \ ,
\end{eqnarray}
and
\begin{eqnarray}
G \gamma_i \eta^* &=& 3G_{ l \bj\bk} \gamma^{ l\bj\bk} 
\gamma_i \eta^* + 3 G_{ lj\bk} \gamma^{ lj\bk} 
\gamma_i \eta^* \nonumber\\
&=& 6 G_{i\bj\bk} \gamma^{\bj\bk} 
\eta^* - 6 G^{\bj}{}_{\bj\bk} \gamma^{\bi\bk} 
\eta^*+ 12 G_{ij}{}^{j} \eta^* \ .
\end{eqnarray}

Expanding the condition~(\ref{Gieta}) in components, we have
\begin{equation}
\partial_i \alpha = 6 Z^{-1/2} G_{ij}{}^{j} = 0\ ,
\end{equation}
where we have made use of the earlier condition~(\ref{Gijk}). Thus,
$\alpha$ is constant (which the Bianchi identities will require to
vanish).  For
$\beta$, we have
\begin{equation}
Z \partial_i \hat\beta_{\bj\bk} = 6 ( G_{i\bj\bk} + G^{\bar l}{}_{\bar l
[\bj} g_{\bk] i})\ . \label{betacon}
\end{equation} 
In order to analyze this condition it is useful to define
\begin{eqnarray}
f_l(z) &=& \frac{1}{2} \hat\epsilon_l{}^{\bj\bk} \hat\beta_{\bj\bk}(z)\ ,
\nonumber\\
G_{il} &=& \frac{1}{2} \hat\epsilon_l{}^{\bj\bk} G_{i\bj\bk}\ ,\quad
G_{\bi\bar l} = \frac{1}{2} \hat\epsilon_{\bar l}{}^{jk} G_{\bi jk}\ .
\end{eqnarray}
Here $\hat \epsilon$ is the numerical $\epsilon$-symbol, with constant
values $0, \pm 1$ regardless of index positions.  Then the
condition~(\ref{betacon}) becomes
\begin{equation}
Z \partial_i f_l = 3(G_{il} + G_{li})\ .
\end{equation}
From the symmetry in $i$ and $l$ it follows that 
\begin{equation}
f_l(z) = \partial_l \phi(z)\ ,\quad \beta_{\bi\bj}(z) = Z^{1/2}
\hat\epsilon_{\bi\bj}{}^l \partial_l \phi(z)
\end{equation}
in terms of a general holomorphic function $\phi(z)$.  Then the full
content of condition~(\ref{Gieta}) is 
\begin{equation}
\partial_i\alpha(z) = 0\ ,\quad G_{il} + G_{li} = \frac{Z}{3} \partial_i
\partial_l \phi(z)\ .  \label{cond2}
\end{equation}
Similarly expanding the final condition~(\ref{Geta}), we find
\begin{eqnarray}
G_{\bi\bj\bk} &=& \frac{1}{6}\hat\epsilon_{\bi\bj\bk} \partial_{\bar l} Z
\,\partial_l \phi\ ,
\nonumber\\
G_{il} - G_{li} &=& -\frac{1}{6} Z^{-1/2} (\alpha\partial_\bj Z
\hat\epsilon^\bj{}_{il} + 4 \partial_{[i}\phi\, \partial_{l]} Z)\ .
\label{cond3}
\end{eqnarray}

In summary, the conditions for ${\cal N }=1$ supersymmetry are
conveniently written by separating $G_{il}$ and $G_{\bi \bar l}$ into
symmetric and antisymmetric parts,
\begin{equation}
G_{il} = S_{il} + A_{il}\ ,\quad 
G_{\bi \bar l} = S_{\bi \bar l} + A_{\bi \bar l}\ .  \label{symant}
\end{equation}
Then $S_{\bi \bar l}$ is completely undetermined, while eqs.~(\ref{Gijk},
\ref{cond2},\ref{cond3}) fix the rest in terms of one constant $\alpha$
and one holomorphic function~$\phi$:
\begin{eqnarray}
G_{ijk} &=& 
A_{\bi\bar l} = 0 \ , \nonumber\\
S_{i l} &=& \frac{Z}{6} \partial_i
\partial_l \phi \ , \nonumber\\
A_{il} &=& -\frac{1}{12} \alpha Z^{-1/2} \partial_\bj Z
\hat\epsilon^\bj{}_{il} -\frac{1}{3} \partial_{[i}\phi\, \partial_{l]} Z \
,
\nonumber\\ 
G_{\bi\bj\bk} &=& \frac{1}{6}\hat\epsilon_{\bi\bj\bk}
\partial_{\bar l} Z\,
\partial_l \phi\ . \label{susysum}
\end{eqnarray}

\subsection{Bianchi identities and equations of motion}

We now impose the Bianchi identity $dG_{(3)} = 0$ on the background. 
Expressed in terms of the fields~(\ref{symant}), these take the form
\begin{eqnarray}
\hat\epsilon^{\bj\bk\bar l} \partial_i G_{\bj\bk\bar l}
&=& 6 \partial_{\bj} G_{ij} \label{bia1}\\
\partial_\bi G_{\bj\bk} - \partial_\bj G_{\bi\bk} 
&=& - \hat\epsilon^{ab}{}_\bk \hat\epsilon^{ci}{}_\bj
\partial_{a} G_{bc}\ ,
\label{bia2}\\
\hat\epsilon^{jkl} \partial_\bi G_{jkl}
&=& 6 \partial_{j} G_{\bi\bj} \ .\label{bia3}
\end{eqnarray}
(We use $a,b,c$ as well as $i,j,k,l$ for holomorphic indices.)
Eq.~(\ref{bia1}) reduces to
\begin{equation}
\partial_i \phi \partial_j \partial_\bj Z = 0\ .
\end{equation}
Thus the Bianchi identity~(\ref{bia1}) holds except at the
locations~(\ref{3den}) of the D3-branes, where it should break down because
these carry 5-brane charges as well in the PS solution.

Eqs.~(\ref{bia2},\ref{bia3}) constrain the remaining components
$G_{\bj\bk}$:
\begin{eqnarray}
\partial_\bi G_{\bj\bk} - \partial_\bj G_{\bi\bk} 
&=& -\frac{1}{24} \alpha Z^{1/2} (\hat\epsilon_{\bk\bi\bj}
w_l w_{\bar l} -
\hat\epsilon^l_{\bi\bj} w_l w_{\bar k} )
- \frac{1}{6} \hat\epsilon^{ab}{}_\bk \hat\epsilon^{c}{}_{\bi\bj}
\partial_{b} \phi\, \partial_{a} \partial_{c} Z\ ,
\label{bia4}\\
\partial_{i} G_{\bi\bj} &=& 0\ .\label{bia5}
\end{eqnarray}
Taking $\partial_i$ of identity~(\ref{bia4}) and using
identity~(\ref{bia5}) gives
\begin{equation}
\partial^2 G_{\bj\bk} =
-\frac{1}{12} \alpha  \partial_i \Bigl[ Z^{1/2} (\hat\epsilon_{\bk\bi\bj}
w_l w_{\bar l} -
\hat\epsilon^l_{\bi\bj} w_l w_{\bar k} ) \Bigr]
- \frac{1}{3} \hat\epsilon^{ab}{}_\bk \hat\epsilon^{ci}{}_\bj
\partial_{b} \partial_{i} \phi\, \partial_{a} \partial_{c} Z\ ;
\label{bia6}
\end{equation}
(note that $\partial^2 = 2 \partial_i \partial_\bi$).
Symmetry in $\bj\bk$ now implies that
\begin{equation}
\alpha = 0\ .
\end{equation}
Inverting eq.~(\ref{bia6})
gives
\begin{equation}
G_{\bj\bk} =
- \frac{1}{3 \partial^2} \hat\epsilon^{ab}{}_\bk \hat\epsilon^{ci}{}_\bj
\partial_{b} \partial_{i} \phi\, \partial_{a} \partial_{c} Z\ .
\end{equation}
To be precise, this final component is not completely determined, because
eq.~(\ref{bia6}) allows us to add any harmonic tensor, subject to the
earlier conditions~$A_{\bj\bk} = \partial_j G_{\bj\bk} = 0$.  The general
solution is then
\begin{equation}
G_{\bj\bk} =
- \frac{1}{3 \partial^2} \hat\epsilon^{ab}{}_\bk \hat\epsilon^{ci}{}_\bj
\partial_{b} \partial_{i} \phi\, \partial_{a} \partial_{c} Z
+ \partial_{\bj}\partial_\bk \psi \label{finsol}
\end{equation}
for $\psi$ any harmonic function.

In summary, the full solution to the supersymmetry conditions and Bianchi
identities is given in eqs.~(\ref{susysum}) (with
$\alpha=0$) and~(\ref{finsol}) in
terms of one holomorphic function $\phi$ and one harmonic function~$\psi$.
In the appendix we give the explicit form for $G_{(3)}$ in the special
case where the cross derivatives of the holomorphic function $\phi$ are
zero. 

Now we will verify that all such solutions satisfy the equations of
motion as well.  To the order we are working, the only nontrivial
equation of motion is that for $G_{(3)}$,
\begin{equation}
d{*}G_{(3)}+ 4i \kappa G_{(3)} \wedge F_{(5)}=0\ .
\label{eom}
\end{equation}
In the 3-brane background, for transverse $G_{(3)}$, this becomes
\begin{equation}
 d[Z^{-1}
G^+_{(3)}]=0\ ,\quad G^+_{(3)} \equiv G_{(3)} + i {*_6} G_{(3)}
\label{eom1}
\end{equation}
where $*_6$ means the dual in the six-dimensional space with respect to the
flat metric $\delta_{mn}$.

Defining for $G^+_{(3)}$ the tensors $G^+_{ij}$, $G^+_{\bi\bj}$, and
their symmetric and antisymmetric parts, in parallel to the earlier
definitions for $G^+_{(3)}$, on finds that
\begin{eqnarray}
G^+_{\bi\bj\bk} &=& A^+_{ij} = S^+_{\bi\bj} = 0\ ,\nonumber\\
G^+_{ijk} &=& 2G_{ijk}\ ,\quad
A^+_{\bi\bj} = 2A_{\bi\bj}\ ,\quad
S^+_{ij} = 2 S_{ij}\ . \label{comdu}
\end{eqnarray}
The only nontrivial component is then
\begin{equation}
S^+_{ij} = \frac{Z}{3} \partial_i \partial_j \phi\ ,
\end{equation}
and the nontrivial equations of motion
\begin{equation}
\partial_{[i} (Z^{-1} S^+_{j]k}) = \partial_{\bi} (Z^{-1} S^+_{ij})
= 0
\end{equation}
are immediately seen to be satisfied.

In appendix A.1 we carry this to second order for the dilaton and axion, and
in appendix A.2 we find the explicit form of $G_{\bi\bj}$ for special $\phi$.

\subsection{Discussion}

On the gauge theory side, the perturbation studied in PS is, in
${\cal N} = 1$ notation, a mass term for the three chiral superfields.  
More generally, any ${\cal N} = 1$ superpotential would preserve one
supersymmetry, and so it is
natural to identify the holomorphic function~$\phi$ with the superpotential.
Let us check that the dimensions are correct, first for the case of pure AdS
space where $Z = R^4/r^4$.  Let $\phi$ be homogeneous of degree $k$,
corresponding to a perturbation of dimension $\Delta = k+1$.
In this case all terms involving
$\phi$ in the solution~(\ref{susysum},\ref{finsol}) for $G_{mnp}$ scale as
$r^{k-6}$, and all terms in the inertial frame $G_{\hat m\hat n\hat p}$ as
$r^{k-3} = r^{\Delta-4}$. This is the correct scaling for the nonnormalizable
solution dual to an operator of dimension~$\Delta$~\cite{witads,GKP},
confirming the interpretation of $\phi$ as dual to a superpotential
perturbation. We have verified that the supergravity solution for $\phi = m
z^i z^i$ reproduces the nonnormalizable solution
\begin{equation}
G_{(3)} = r^{-4} (T_{(3)} - 4V_{(3)}/3)
\end{equation}
where (in the notation of PS) the nonzero components of $T$ are
\begin{equation}
T_{i\bj\bk} = m \hat \epsilon_{i\bj\bk}\ ,
\end{equation}
and $V_{(3)}/3$ is $T_{(3)}$ projecting out components orthogonal to $x^m$.

The PS solutions also have a normalizable part; with its
inclusion the solution must still be supersymmetric.  The solutions we have
found here have no independent normalizable part; in particular, we will see
that $\psi$ corresponds to higher dimensional operators.  Rather, the
normalizable part is already determined in terms of $\phi$ and $Z$.  With
expanded 5-branes $Z$ has terms subleading in $1/r$, and through the
solution~(\ref{susysum},\ref{finsol}) these generate the normalizable part of
$G_{(3)}$.  In the appendix we verify that the component~$G_{\bi\bj\bk}$
obtained here is in agreement with PS.

The harmonic function $\psi$ produces solutions with the same $SO(6)$ quantum
numbers as $\phi$ and with dimension $\Delta'$ greater by 4.  
This follows from eq.~(\ref{finsol}), where the two terms have the same
net number of derivatives but the first has an extra factor of $r^{-4}$
asymptotically from $Z$. Both branches
appear in table~2 of ref.~\cite{krv} (see also ref.~\cite{gunmar}).  Just as
the superpotential perturbations correspond to operators of the form
$\lambda^i \lambda^j \partial_i\partial_j \phi$, the solutions given by
$\psi$ have the dimensions and $SO(6)$ quantum numbers of the operators
$F^2 \bar\lambda^i \bar\lambda^j \partial_\bi\partial_\bj \psi$.  We have not
fully understood from the field theory side why the latter are parameterized
by a harmonic rather than holomorphic function.

\sect{A class of exact solutions}

In this section we note an interesting class of exact solutions with
$G_{(3)}$ flux.  We begin with a Calabi-Yau background,
\begin{equation}
\widetilde{ds^2} = \eta_{\mu\nu} dx^\mu dx^\nu + \widetilde{ds_K^2}\ ,
\end{equation}
with $\widetilde{ds_K^2}$ is a Ricci-flat metric on the transverse space
$K$.   The dilaton-axion field is constant,
\begin{equation}
\tau = \frac{\theta}{2\pi} + \frac{i}{g}\ ,
\end{equation}
and the other IIB supergravity fields vanish,
\begin{equation}
F_{(3)} = H_{(3)} =  F_{(5)} = 0\ .
\end{equation}
The dilatino variation vanishes trivially, while the gravitino variation
vanishes for
\begin{equation}
\partial_\mu \tilde\varepsilon = \tilde D_m \tilde\varepsilon = 0
\end{equation}
Thus there are two $D=4$ supersymmetries (from the real and imaginary parts
of $\varepsilon$) for each covariantly constant spinor on $K$.

Now introduce a warp factor $Z(x^m)$,
\begin{equation}
{ds^2} = Z^{-1/2} \eta_{\mu\nu} dx^\mu dx^\nu + Z^{1/2} 
\widetilde {ds_K^2} 
\end{equation}
and a five-form flux (\ref{F5}),
\begin{equation}
 F_{\it 5} = d\chi_{\it 4} + * d\chi_{\it 4}\ ,\quad
\chi_{\it 4} = \frac{1}{4\kappa Z}  dx^0 {\wedge} dx^1 {\wedge} dx^2
{\wedge}  dx^3\ ,
\end{equation}
with constant dilaton-axion and vanishing three-form fluxes.
The dilatino variation is zero, while
\begin{eqnarray}
\kappa\delta \psi_\mu &=& \partial_\mu \varepsilon - \frac{1}{8}
\gamma_\mu
\gamma_w (1 - \Gamma^4) \varepsilon \ ,\nonumber
\\
\kappa\delta \psi_m &=& \tilde D_m \varepsilon + \frac{1}{8}
\varepsilon
w_m - \frac{1}{8} \gamma_w \gamma_m (1 - \Gamma^4) 
\varepsilon \ .
\end{eqnarray}
There is then an unbroken supersymmetry~\cite{kehag}
\begin{equation}
\varepsilon = Z^{-1/8} \tilde \varepsilon \label{vareps}
\end{equation}
for each covariantly constant spinor of chirality $\Gamma^4\tilde
\varepsilon = \tilde \varepsilon$.
This is the familiar multi-three-brane metric: it is a sourceless solution to
the equations of motion
for $Z$ a harmonic  function of the transverse coordinates, and more
generally is a solution with D3-brane sources
\begin{equation}
-\tilde \nabla^2_K Z = (2\pi)^4 \alpha'^2 g \rho_3(x^m)\ .
\end{equation}

Now we construct a solution with nonzero $G_{(3)}$.  We need on $K$ a 3-form
$\omega_{(3)}$ which is both closed and divergenceless,
\begin{equation}
d \omega_{(3)} = d {*_{K}} \omega_{(3)} = 0\ .
\end{equation}
A harmonic 3-form on a compact manifold, or with sufficiently rapid falloff
on a noncompact manifold, will have this property.  By forming linear
combinations (and taking a complex conjugate if needed) we may assume that
\begin{equation}
{*_{K}} \omega_{(3)} = i \omega_{(3)}  \label{imdu}
\end{equation}
Then $G_{(3)} = C \omega_{(3)}$ solves the equations of motion, where the
metric, dilaton-axion, and 5-form are still of black 3-brane form.  The 
$G_{(3)}$ equation of motion~(\ref{eom1}) is trivial.  The field equation for
the metric, at constant $\tau$, is
\begin{eqnarray}
R_{MN} - \frac{\kappa^2}{6}
 F_{MPQRS}  F_N{}^{PQRS} &=& 
\frac{\kappa^2}{4} G_{(M|PQ} G_{|N)}{}^{PQ*} 
- \frac{\kappa^2}{48} g_{MN} G_{PQR} G^{PQR*}\ . \label{einst}
\end{eqnarray}
For the black 3-brane Ansatz, the left-hand side is
\begin{eqnarray}
{ L}_{\mu\nu} &=& \eta_{\mu\nu}\frac{1}{4Z^2} \tilde\nabla^2 Z \ ,\quad
{ L}_{mn} = - \tilde g_{mn} \frac{1}{4Z} \tilde\nabla^2 Z\ .
\end{eqnarray}
On the right-hand side, the condition~(\ref{imdu}) implies that
\begin{eqnarray}
G_{mpq} G_{n}{}^{pq*} &=& \frac{1}{36} \epsilon_{mpq}{}^{rst}
\epsilon_{n}{}^{pquvw} G_{rst} G_{uvw}^* \nonumber\\
&=& \frac{1}{3} g_{mn} G_{rst} G^{rst*} - G_{npq} G_{m}{}^{pq*}\ ,
\end{eqnarray}
and so
\begin{equation}
G_{(m|pq}G_{|n)}{}^{pq*} = \frac{1}{6} g_{mn} G_{pqr} G^{pqr*}\ .
\end{equation}
The field equation is then satisfied for
\begin{equation}
-\tilde\nabla^2 Z = (2\pi)^4 \alpha'^2 g \rho_3(x^m) + \frac{\kappa^2}{12} 
G_{pqr} G^{\widetilde{pqr}*}\ . \label{source}
\end{equation}
Note that it is the original, tilded, metric on $K$ that appears here.  The
Bianchi identity,
\begin{equation}
d F_{(5)} = -4i \kappa G_{(3)} \wedge G_{(3)}^*\ ,
\end{equation}
is satisfied under the same condition~(\ref{source}).

Not all these solutions are supersymmetric, but they become so if we impose
the additional condition that $\omega_{(3)}$ contain only $(2,1)$ and $(1,2)$
components under the complex structure defined by the supersymmetry of the
Calabi-Yau manifold $K$.  In other words,
\begin{equation}
\omega_{(3,0)} = \omega_{(0,3)} = 0\ . \label{30}
\end{equation}
To see this, note first that the self-duality condition~(\ref{imdu}) implies in
the notation of eq.~(\ref{comdu}) that $\omega^+ = 0$ or
\begin{equation}
\omega_{ijk} = \alpha_{\bi\bj} = \sigma_{ij}
= 0\ ,
\end{equation}
where $\sigma$ and $\alpha$ are the symmetric and antisymmetric parts of
$\omega$, defined by analogy to the earlier $S$ and $A$.
The (0,3) condition~(\ref{30}) implies that also $\omega_{\bi\bj\bk} =
0$. Finally, $\omega_{ij}{}^j$ must vanish, else it would be a harmonic (1,0)
form, which does not exist on a Calabi-Yau manifold.  This is equivalent to 
$\alpha_{ij} = 0$.  It follows that the only nontrivial component of
$\omega_{(3)}$ is $\sigma_{\bi\bj}$.
Now we must consider the supersymmetry conditions, treating $\varepsilon$
exactly.  We claim that the fermionic fields remain invariant for the same
spinor~(\ref{vareps}) as in the absence of $G_{(3)}$.  The terms that do not
involve $G_{(3)}$ in the variations~(\ref{dilatino},\ref{gravitino}) already
vanish, so those that involve $G_{(3)}$ must vanish separately.  By a
calculation directly parallel to that which led to eq.~(\ref{susysum}), 
one sees that the nonzero components $\sigma_{\bi\bj}$ do not appear in the
variations, which therefore vanish.

Note the structure of this solution, with its strong resemblance to the
D3-brane Higgs branch solution.
The flux
$G_{(3)} \wedge
G_{(3)}^*$ behaves like an additional D3-brane density, but without the
moduli of D3-branes --- perhaps one can think of this flux as a sort of frozen
density of D3-branes.  In fact, this freezing seems to be a manifestation of
confinement: taking $K$ to be the deformed conifold, this is precisely the
recent KS solution~\cite{klebstrass}.  For the conifold
itself, we obtain the solution of Klebanov and
Tseytlin~\cite{klebtsey}, which is singular because the integrated
D3-brane density diverges; the resolved conifold would lead to a similar
singular solution.  For
${\bf R}^2
\times {\bf R}^4 / {\bf Z}_2$ one obtains an ${\cal N}=2$ analog of the
KS solution~\cite{joemm}.

The solution found in this section resembles the solution found by
Becker and Becker for M theory on a Calabi-Yau four-fold~\cite{beck^2}.
In fact, it is a special case, if one takes the four-fold to be a
three-fold times $T^2$, and then takes the $T$-dual on $T^2$ as in
refs.~\cite{GVW,DGS} to obtain a IIB solution.

\sect{Conclusions}

We have verified the supersymmetry of the PS solution to first
order in the perturbation, and of its
generalization to an arbitrary ${\cal N} = 1$ superpotential, and we have
shown that this condition together with the Bianchi identity determines
the solution.  The ${\cal N} = 1$ conditions may be a useful method to
find the exact solution, and so describe physics that is outside the
approximation used in PS.

We have also found an interesting exact solution, which includes the
KS solution but not the PS solution --- the three-form flux in the
latter case is not of the form ${*_{K}} G_{(3)} = i G_{(3)}$, and the
dilaton is not constant.  It would be useful to find a generalization which
includes both solutions (and also the solution~\cite{maldanun}), and
so obtain a more universal understanding of the supergravity duals of
confining theories.  The more general Becker--Becker
solutions~\cite{beck^2,GVW,DGS} may be useful here.\footnote
{The relevance of these solutions to Randall--Sundrum compactification
has recently been discussed in refs.~\cite{recent}.}
It is a further useful direction to incorporate these noncompact
solutions as local regions in a compactified space (as
in~\cite{Verlinde}), and so produce Randall--Sundrum type
compactifications with large warp factors and four-dimensional
gravity.

\subsection*{Acknowledgments}

This work was supported by National Science
Foundation grants PHY99-07949 and PHY97-22022.

\appendix
\sect{Appendix}

\subsection{Second order dilaton and axion}

At second order in the expansion there is a nonconstant dilaton and
axion and a correction to the metric and $F_{(5)}$.  These have recently
been obtained directly from the equations of motion by Freedman and
Minahan~\cite{freemin}, who also considered the finite temperature case.
Here we will simply verify that the
supersymmetry equations determine the second order (zero
temperature) dilaton and axion.  The second order dilatino
variation involves only these second order corrections,
\begin{equation}
\frac{i}{\kappa}\gamma^{M}P_{M}\varepsilon_0^{*}-
\frac{i}{24}\gamma^{MNP}G_{MNP} \varepsilon_1=0\ .
\label{dialtino2}
\end{equation}
Inserting the first order solution, this becomes
\begin{equation}
P_{\bi}\gamma^{\bi} \eta^{*} = 
\frac{\kappa^2}{24^2 S^2} G \tilde \gamma G \eta^{*} \nonumber\\
= \frac{\kappa^2}{24} G_{\bi\bj} \partial_j \phi \gamma^{\bi} \eta^{*}
\ .
\end{equation}
Expanding $B = B_0 + \delta B$, this is
\begin{equation}
f_0^2\partial_\bi \delta B = \frac{\kappa^2}{24} G_{\bi\bj} \partial_j
\phi\ ,
\label{dbitau}
\end{equation}
which is integrable by eq.~(\ref{bia4}):
\begin{equation}
\delta B =  \frac{\kappa^2}{12 f_0^2}\frac{1}{ \partial^2} G_{\bi\bj}
\partial_i\partial_j \phi\ .\label{deltau}
\end{equation}
The condition~(\ref{dbitau}) allows an arbitrary holomorphic piece
$F(z)$ in $\delta B$.  This corresponds to an additional ${\cal N} =
1$ interaction
\begin{equation}
\int d^2\theta\,F(\Phi) W_\alpha W^\alpha\ .
\end{equation}

\subsection{Particular solution}

In eqs.~(\ref{finsol}) and~(\ref{deltau}) we have given the solutions in
terms of the Green function $\partial^{-2}$.  Here we note that for
$Z = R^4/r^4$ and $\phi$ of
the form
\begin{equation}
\phi = \sum_{i=1}^3 f_i(z^i)
\end{equation}
(which includes the mass term $\frac{1}{2} m z^i z^i$ as a special case)
we can give a closed form for each.  Specifically, 
\begin{equation}
G_{\bar{1}\bar{1}}=\frac{2 R^4}{3r^6}\Biggl[\bar{z}^2 \bar{z}^2
\frac{\partial_3
\phi}{z^3}+ \bar{z}^3 \bar{z}^3
\frac{\partial_2
\phi}{z^2}\Biggr]+\partial_{\bar{1}}\partial_{\bar{1}}\psi
\label{g11}
\end{equation}
(and the same permuted for the other diagonal terms) and
\begin{equation}
G_{\bar{1}\bar{2}}=\frac{2 R^4}{3r^6}
 \bar{z}^1\bar{z}^2 \frac{\partial_3
\phi}{z^3} +\partial_{\bar{1}}\partial_{\bar{2}}\psi
\label{g12}
\end{equation}
and permutations for the off-diagonal terms.
For the second order dilaton and axion, we obtain:
\begin{equation}
\delta B =
-\frac{\kappa^2 R^4}{144 f_0^2 r^4}\Biggl(\bar{z}^1\bar{z}^1
\frac{\partial_2\phi
\partial_3\phi}{z^2 z^3}+\bar{z}^2\bar{z}^2 \frac{\partial_1\phi
\partial_3\phi}{z^1 z^3}+\bar{z}^3\bar{z}^3 \frac{\partial_1\phi
\partial_2\phi}{z^1 z^2}\Biggr) +
\frac{\kappa^2}{24 f_0^2}\partial_i\phi\partial_{\={\i}}\psi + H(z)\ ,
\label{dilat2}
\end{equation}
where $H$ is any holomorphic function.

\subsection{Comparison to PS}

In this appendix we compare our solution to that of PS, in particular to
verify that the normalizable part arises as argued in section~3.3.  We focus
on the solution with a single D5-sphere.  The PS solution was
\begin{equation}
G_{(3)} = {*_6} d\omega_{(2)} + i d\omega_{(2)} + d \eta_{(2)}\ ,
\label{gsol}
\end{equation}
where $\eta_{(2)}$ is the background field, while the $\omega_{(2)}$ terms are
from the brane source.

The potentials are
\begin{eqnarray}
\omega_{(2)} &=&  -\frac{\alpha'}{4w^3}
\epsilon_{ijk} w^i dw^j \wedge dw^k
\Biggl( -\ln\frac{A}{B} + \frac{2r_0 w}{A} + \frac{2 r_0 w}{B} \Biggr)\ ,
\nonumber\\
\eta_{(2)} &=& \frac{\alpha'}{2 \sqrt{2} m w^3} T_{mnp} dx^n \wedge dx^p
\nonumber\\
&&\qquad\times
\biggl( - w w_{,m} \ln\frac{A}{B} + 2 (w+r_0) w^2 w_{,m} \biggl[
\frac{1}{A}+\frac{1}{B} \biggr] + 2 w^2 y y_{,m} \biggl[
\frac{1}{A}+\frac{1}{B} \biggr] \Biggr)\ ,
\label{onsol}
\end{eqnarray}
where $r_0 = \pi\alpha' m N$ and
\begin{eqnarray}
y^i &=& \frac{z^i + \bar z^i}{\sqrt 2}\ ,\quad
w^i = i\frac{z^i - \bar z^i}{\sqrt 2}\ ,\nonumber\\
A &=& y^2 + (w+r_0)^2 \ ,\quad B = y^2 + (w-r_0)^2 \ .
\end{eqnarray}

This does indeed agree with the result in section~3.  Consider for example
the component $G_{\bar 1 \bar 2 \bar 3}$, for which the earlier result was
\begin{equation}
G_{\bar 1 \bar 2 \bar 3} = \frac{1}{6} \partial_{\bar l} Z \partial_l \phi\ ,
\end{equation}
corresponding to the potential ($G_{(3)} = d\Lambda_{(2)}$)
\begin{equation}
\Lambda_{\bj\bk} = \frac{1}{6} \hat\epsilon^i{}_{\bj\bk} Z \partial_i \phi
\label{present}\ .
\end{equation}
For the $\bar 1 \bar 2 \bar 3$ component, ${*_6} d\omega_{\it 2} = i
d\omega_{\it 2}$ and so the PS solution takes the form
\begin{equation}
G_{\bar 1 \bar 2 \bar 3}= d(\eta + 2i\omega)_{\bar 1 \bar 2
\bar 3}.
\end{equation}
For the solution~(\ref{onsol}) we find
\begin{equation}
(\eta + 2i\omega )_{\bj\bk} = 
\frac{m N \alpha'^2 \sqrt 2}{AB} z^i \hat\epsilon^i{}_{\bj\bk}\ .
\label{psres}
\end{equation}

We should note the translation between Schwarz's
conventions, used here, and the conventions in PS:
\begin{equation}
F_{(5)}^{\rm PS}  = \frac{4\kappa}{g} F_{(5)}^{\rm S}\ ,\quad
G_{(3)}^{\rm S}  = \frac{\kappa}{g} G_{(3)}^{\rm S}\ .
\end{equation}
(The general normalization for $G_{(3)}^{\rm S}$ beyond linear order is
more complicated; also we have assumed for convenience that $\theta = 0$.)
Noting also that $Z = R^4/AB$, the present~(\ref{present}) and
PS~(\ref{psres}) results agree for $\phi = 3\sqrt{2} (g/\kappa) m z^i z^i$,
which is indeed the superpotential up to a numerical constant.


\begin{thebibliography}{10}
\baselineskip=12pt

\bibitem{Maldacena}
J. Maldacena, Adv.\ Theor.\ Math.\ Phys.\ {\bf 2}, 231 (1998)
[hep-th/9711200].

\bibitem{RS1}L.~Randall and R.~Sundrum, Phys. Rev. Lett. {\bf 83}, 3370 (1999)
[hep-th/9905221]

\bibitem{RS2}L.~Randall and R.~Sundrum, Phys. Rev. Lett. {\bf 83}, 4690 (1999)
[hep-th/9906064]

\bibitem{Verlinde} H.Verlinde, Nucl. Phys. {\bf B580}, 264 (2000) [hep-
th/9906182]

\bibitem{witads}
E.~Witten,
Adv.\ Theor.\ Math.\ Phys.\  {\bf 2}, 253 (1998)
[hep-th/9802150].

\bibitem{GKP}
S.~S.~Gubser, I.~R.~Klebanov and A.~M.~Polyakov,
Phys.\ Lett.\  {\bf B428}, 105 (1998)
[hep-th/9802109].

\bibitem{VafaWitten}C.Vafa and E.Witten, Nucl. Phys. {\bf B431}, 3 (1994)
[hep-th/9408074].

\bibitem{DonagiWitten}R.Donagi and E.Witten, Nucl. Phys. {\bf B460}, 299
(199) [hep-th/9510101].

\bibitem{Strassler}M.Strassler, Prog. Theor. Phys. Suppl. {\bf 131}, 439
(1998) [hep-lat/9803009].

\bibitem{Dorey}N. Dorey, JHEP {\bf 9907}, 021 (1999) [hep-th/9906011].

\bibitem{DoreyKumar}N. Dorey and S. Kumar, JHEP {\bf 0002}, 006 (2000)
[hep-th/0001103].

\bibitem{gppz}
L.~Girardello,\ M.~Petrini,\ M.~Porrati and A.~Zaffaroni,
[hep-th/9909047].

\bibitem{PS}
J.~Polchinski and M.~Strassler, hep-th/0003136.

\bibitem{kacsil}
S.~Kachru and E.~Silverstein,
Phys.\ Rev.\ Lett.\  {\bf 80}, 4855 (1998)
[hep-th/9802183].

\bibitem{kehag}
A.~Kehagias,
Phys.\ Lett.\  {\bf B435}, 337 (1998)
[hep-th/9805131].

\bibitem{klebwit}
I.~R.~Klebanov and E.~Witten,
Nucl.\ Phys.\  {\bf B536}, 199 (1998)
[hep-th/9807080].

\bibitem{morpless}
D.~R.~Morrison and M.~R.~Plesser,
Adv.\ Theor.\ Math.\ Phys.\  {\bf 3}, 1 (1999)
[hep-th/9810201].

\bibitem{gubkleb}
S.~S.~Gubser and I.~R.~Klebanov,
Phys.\ Rev.\  {\bf D58}, 125025 (1998)
[hep-th/9808075].

\bibitem{klebnek}
I.~R.~Klebanov and N.~A.~Nekrasov,
Nucl.\ Phys.\  {\bf B574}, 263 (2000)
[hep-th/9911096].

\bibitem{klebtsey}
I.~R.~Klebanov and A.~A.~Tseytlin,
Nucl.\ Phys.\  {\bf B578}, 123 (2000)
[hep-th/0002159].

\bibitem{klebstrass}
I.~R.~Klebanov and M.~J.~Strassler,
JHEP {\bf 0008}, 052 (2000)
[hep-th/0007191].

\bibitem{joemm}
J.~Polchinski, Proceedings of Strings '00, in preparation.

\bibitem{beck^2}
K.~Becker and M.~Becker,
Nucl.\ Phys.\  {\bf B477} (1996) 155
[hep-th/9605053].

\bibitem{GVW}
S.~Gukov, C.~Vafa and E.~Witten,
Nucl.\ Phys.\  {\bf B584}, 69 (2000)
[hep-th/9906070].

\bibitem{DGS}
K.~Dasgupta, G.~Rajesh and S.~Sethi,
JHEP {\bf 9908}, 023 (1999)
[hep-th/9908088].

\bibitem{Schwartz}
J. Schwarz and P. West, Phys.\ Lett.\ {\bf B126}, 301 (1983);\\
J. Schwarz, Nucl.\ Phys.\ {\bf B226}, 269 (1983);\\
P. Howe and P. West, Nucl.\ Phys.\
{\bf B238}, 181 (1984).

\bibitem{candelas}
P.~Candelas,
Nucl.\ Phys.\  {\bf B256}, 385 (1985).

\bibitem{dewit}
B.~de Wit, D.~J.~Smit and N.~D.~Hari Dass,
Nucl.\ Phys.\  {\bf B283}, 165 (1987).

\bibitem{krv}
H.~J.~Kim, L.~J.~Romans and P.~van Nieuwenhuizen,
Phys.\ Rev.\  {\bf D32}, 389 (1985).

\bibitem{gunmar}
M.~Gunaydin and N.~Marcus,
Class.\ Quant.\ Grav.\  {\bf 2}, L11 (1985).

\bibitem{maldanun}
A.~H.~Chamseddine and M.~S.~Volkov,
  Phys.\ Rev.\ Lett.\  {\bf 79}, 3343 (1997)
  [arXiv:hep-th/9707176];\\
  A.~H.~Chamseddine and M.~S.~Volkov,
  Phys.\ Rev.\  D {\bf 57}, 6242 (1998)
  [arXiv:hep-th/9711181];\\
J.~M.~Maldacena and C.~Nunez,
hep-th/0008001.

\bibitem{recent}
K.~Behrndt and S.~Gukov,
Nucl.\ Phys.\  {\bf B580}, 225 (2000)
[hep-th/0001082];\\
C.~S.~Chan, P.~L.~Paul and H.~Verlinde,
Nucl.\ Phys.\  {\bf B581}, 156 (2000)
[hep-th/0003236];\\
B.~R.~Greene, K.~Schalm and G.~Shiu,
Nucl.\ Phys.\  {\bf B584}, 480 (2000)
[hep-th/0004103].

\bibitem{freemin}
D.~Z.~Freedman and J.~A.~Minahan,
hep-th/0007250.

\end{thebibliography}
\end{document}